# AI-Powered Spearphishing Cyber Attacks: Fact or Fiction?

Matthew **Kemp**[a], Harsha **Kalutarage**[a] and M. Omar **Al-Kadri**[b]

[a]*School of Computing, Robert Gordon University, United Kingdom*
[b]*School of Computing and Digital Technology, Birmingham City University, United Kingdom*

## ABSTRACT

Due to society's continuing technological advance, the capabilities of machine learning-based artificial intelligence systems continue to expand and influence a wider degree of topics. Alongside this expansion of technology, there is a growing number of individuals willing to misuse these systems to defraud and mislead others. Deepfake technology, a set of deep learning algorithms that are capable of replacing the likeness or voice of one individual with another with alarming accuracy, is one of these technologies. This paper investigates the threat posed by malicious use of this technology, particularly in the form of spearphishing attacks. It uses deepfake technology to create spearphishing-like attack scenarios and validate them against average individuals. Experimental results show that 66% of participants failed to identify AI created audio as fake while 43% failed to identify such videos as fake, confirming the growing fear of threats posed by the use of these technologies by cybercriminals.

## 1. Introduction

It can be confidently stated that Artificial Intelligence (AI) and Neural Network (NN) technologies have been deployed to enhance a vast myriad of professions and industries. For example; Health, Marketing and Robotics. However, less commonly considered is the opportunity presented to those who are willing to creatively misuse these tools; namely cybercriminals. There is a growing fear among security professionals of the threat presented by cybercriminals willing to use AI to augment their attack strategies. A recent report by the Royal United Services Institute for Defence and Security Studies (RUSI) states that "Malicious actors will undoubtedly seek to use AI to attack the UK, and it is likely that the most capable hostile state actors, which are not bound by an equivalent legal framework, are developing or have developed offensive AI-enabled capabilities" [16]. It warns that while advances in AI create new opportunities to increase cybersecurity, it also creates potential new challenges, including the risk that attackers could use the same technology for their benefit. This paper aims to evaluate the threat posed by this type of attack, namely Spearphishing attacks, through the creation and evaluation of artificial media using deep learning technologies.

The contribution of this paper begins with a comprehensive review of the literature. As this is an emerging threat, the established academic literature on the use of deep learning technologies for phishing attacks is scarce. Therefore, we cover both exciting scientific literature and some carefully selected Grey literature in this area. Then we use deep learning-enabled technologies to mimic spearphishing like attack scenarios and validate them against average individuals. Experimental results showed that 66% of the participants were unable to recognise artificially generated audio as fake, while 43% could not identify artificially generated video as fake, regardless of the low-spec computer facilities used in these experiments. The authenticity of artificially created videos could be further enhanced through the use of high-spec computing facilities, which are easily accessible for most capable hostile state threat actors.

The remainder of the paper is structured as follows: Section 2 provides information related to the background associated with this research. Section 3 presents a comprehensive review of related literature. It contains details of reported phishing attacks using deep learning technologies and a review of existing detection methods. Section 4 details the experimental setting, including the software utilised to create the deepfake media evaluated in our study. Section 5 discusses the experimental results and observations. Finally, Section 6 concludes the paper and proposes potential work that could be undertaken in the future.

## 2. Background

This section provides a background on deepfake technology and the associated contemporary techniques. This will be accompanied by a brief explanation of the inception of phishing and its efficacy in the UK.

### 2.1. Deepfake Technology

The name 'Deepfake' is a portmanteau of "deep learning" and "fake" and was first coined by a user on Reddit who had used machine learning algorithms to edit celebrities into explicit video material. Deepfakes are based on a machine learning framework named Generative Adversarial Networks (GANs) which create fake videos by superimposing a person's facial features upon another's using existing imagery and footage of the subject as reference. In the context of Deepfake creation, this allows for the GAN model to repeatedly identify any weak points of a generated product and correct them; iteratively enhancing the realism of the final output. This process can be applied to both video and audio. Though the term 'deepfake' was created regarding the malicious uses of deep learning algorithms; deepfakes can be used for benign and useful applications. For example; Zhu et al.[56] note that deepfakes could be used to obfus-

ORCID(s):

cate patients' facial features in medical imagery. Therefore, protecting their privacy. Furthermore, Forbes[27] suggests that deepfakes may be utilised to recreate the likenesses of significant people from history.

The Centre for Data Ethics and Innovation[22] identifies four primary forms of deepfake. These are; face replacement, face re-enactment, face generation and speech synthesis.

**Face Replacement:** Face replacement can be considered the most familiar form of deepfake and typically involves various samples of an individual alongside an existing video clip the user wishes to add the individual into. These are most commonly created using applications such as 'DeepFaceLab'[28]. The algorithm then identifies the boundaries of all faces within the supplied media in addition to key features such as the eyes, bridge of the nose and facial structure. Once these are mapped, the model then superimposes the sample individual's face over the existing video. This allows for the replacement of characters in existing media without the use of existing facial motion capture techniques commonly used in the film industry[36].

**Face Re-enactment:** Face re-enactment operates with the same methodology as face replacement but instead of replacing the individual within the media, it is used to alter their facial features, particularly the mouth so that they appear to be saying something other than what was originally said. This technique is greatly enhanced when deployed alongside genuine or deepfake audio of the individual.

**Face Generation:** Face generation makes use of technology such as the 'StyleGAN' architecture which allows for specific characteristics to be isolated and modified[31]. These are deployed in combination with sample imagery to generate entirely new faces using the facial details and structures from the supplied media alongside the custom characteristics designated. This technique was developed in conjunction with NVidia, a technology and hardware company; with the aim to generate more realistic characters in movies and video games.

**Speech Synthesis:** The final deepfake technique; speech synthesis uses samples of an individual's voice which are then broken down and isolated as a Mel-spectrographic representation[29]. Sections of these representations may then be manipulated and utilised alongside a textual input to generate the script as spoken by the sampled individual.

## 2.2. Phishing

The act of phishing is a form of social engineering attack in which a malicious actor attempts to gain access to another's information or account details by misleading their victim into believing they are interacting with a legitimate authority or organisation[45]. The term 'phishing' was first used on January the 2nd 1996[43] on a Usenet forum named AOHell by Koceilah Rekouche who at the time; used the online handle "da Chronic"[46]. Rekouche explains that he used the term to describe the method his friend utilised to steal credit card information from new AOL users; normally by posing as 'billing' employees asking for account verification. Since its inception, the use of phishing has grown rapidly with 32% of UK organisations reporting phishing style attacks in 2016[32] further increasing to 86% in 2020[25]. Furthermore, the techniques have since been refined to form 'spearphishing'; the act of targeting one specific individual with a highly specialised attack and 'whaling' where the attacker intentionally targets high-level members of an organisation such as senior executives by impersonating their superiors and requesting the transferal of funds or information. The National Cyber Security Centre (NCSC) report mentions that email domains and templates are commonly recreated with slight changes to maintain realism[21].

## 3. Related Work

This section aims to highlight previous instances of video and audio deepfakes used to steal information or mislead. These attacks and their effectiveness will then be compared and contrasted. In addition to this; due to the scarcity of existing examples of deepfake phishing attacks; highly sophisticated spearphishing attacks that could benefit from deepfake technology will also be analysed.

Furthermore, additional threats posed by deepfake media will also be presented alongside a review of detection methods.

### 3.1. Spearphishing Attacks

It should be noted that the attackers were ultimately apprehended due to the suspicious nature of their requests which shared similarities with previous scam attempts.

The Guardian[3] reported an incident in which a group of attackers orchestrated a scheme targeting chief executive officers (CEOs) of French stock exchange companies. Posing as representatives of a government agency, they falsely claimed to be seeking funds to pay the ransom for French hostages in Syria and Mali. Such attacks are referred to as 'whaling', a form of spearphishing that specifically targets high-profile individuals such as CEOs or government officials. The attackers went to great lengths to make their deception convincing. They created a highly accurate replica of a government official's office and employed a makeup professional to disguise one individual as the official, while others acted as their staff. Using this elaborate setup, they contacted company leaders who had previously responded to their requests and negotiated via Skype for a total sum of €8 million. Ultimately, the perpetrators were apprehended due to the suspicious nature of their demands, which bore similarities to previous scam attempts.

Though the above example demonstrates that spearphishing attacks impersonating high-level individuals do not explicitly require deepfake technology, it does establish that if an attacker is careful to avoid common tactics utilised by scammers they can potentially deceive their victims. Moreover, if an attack such as this was enhanced through the use of an audio synthesis algorithm to emulate the foreign minister's voice an even greater degree of realism could be achieved.

Considering this case; it can be suggested that regardless of how realistic an impersonation might be, many professionals are familiar with the tactics used by fraudulent organisations. Furthermore, though it is not noted in the article, it can be assumed that if an attacker chooses a high-level individual to impersonate; they greatly increase the likelihood that a potential victim may decide to contact the actual person for verification if circumstances appear particularly suspicious or unusual.

A second, more advanced case was reported by The Wall Street Journal[10], involving the first documented deepfake-enabled spearphishing attack. An attacker targeted a UK-based energy company by using artificial intelligence to replicate the voice of a senior executive (CEO) from the company's parent organization. The impersonated voice instructed the victim to transfer €220,000 to a specified account, claiming it was for an urgent payment to a supplier. The victim later noted that they were convinced they were speaking with their superior, as the voice mimicked the individual's distinct accent and vocal nuances[10].

Through these observations, it can be noted that a key component of a convincing deepfake attack is careful attention to detail. Accents, idiosyncrasies and favoured vernacular should all be considered when both selecting an individual to impersonate and synthesising an audio sample. Moreover, in an interview with the Wall Street Journal, Mr Filar; the director of data science at a company named 'Endgame' theorised; "imagine a video call with [a CEO's] voice, the facial expressions you're familiar with, then you wouldn't have any doubts at all"[10]. This statement can be easily related to the case in [3] previously analysed, as the accurate representation of the government official is ultimately what convinced the victims to part with company funds. Furthermore, if an individual without prior warning both sees and hears someone they know, they are highly unlikely to assume something is amiss.

Additionally, it can be observed in both attack cases in [10] and [3], that the fraud was identified, when the attackers proceeded to make a second request for additional finances using a different phone number; both of which are characteristics commonly associated with fraudulent activities. Ultimately, there are very few examples of high-level spearphishing attacks that deploy deepfake technology. Regardless, the opportunity for integration into existing phishing tactics is available.

## 3.2. Deepfakes Outside of Phishing

Deepfakes can have alarming impacts outside the application of phishing attacks. This section reviews such attempts in the literature, non-phishing related deepfake threats.

AI Business[5] states that "deepfakes also have significant potential to enhance market manipulation attacks" and notes that highly visible individuals such as prominent CEOs of major technology companies make optimal targets, as their unpredictable behaviour makes damaging actions more believable. For example, on the 1st of May 2020, the CEO of a leading electric vehicle manufacturer tweeted that their company's stock price was "too high imo"[6]. This resulted in the company's stock prices dropping by 13%.

An additional consideration when selecting an individual to impersonate is whether there is a large amount of existing audio or video data that may be utilized to generate a deepfake. In this case, the CEO in question has appeared on podcasts and made numerous public presentations. However, a more secretive executive may feature in far fewer media outlets, reducing the data available for deepfake generation.

In contrast to AI businesses'[5] suggestion of selecting an individual known for their brashness. An alternative may be for an attacker to select a figure largely unknown to the public. This would enable an attacker to exercise more freedom when deciding how to present their deepfake individual.

An example of the devastating impact fake information can have on financial sectors was highlighted by a tweet from a prominent news organization's Twitter account, which had been hacked by a black-hat group known as the 'Syrian Electronic Army'[35]. The tweet falsely claimed that explosions had occurred at a significant government building and that a high-ranking official had been injured. Once reported by the media, this misinformation caused a loss of $136 billion in market capital.

Consequentially, if an organisation creates and coordinates a feasible chain of events such as a deepfake audio clip of a CEO 'secretly' making offensive remarks followed by then falsifying tweets from the CEO defending their actions they can severely impact the value of their target's stocks. If this attack is successful, the attacker - or those who hired them - may then invest in or sell the affected stocks. Furthermore, it would likely be difficult to link the two occurrences together if the amount of money traded is of a sensible size, making investigations more challenging.

Deepfake media has the potential to be deployed in other forms of cybercrime and can have a devastating impact on society[30]. Wojewidka[53] suggests that as biometric security systems such as facial recognition and voice authentication become more widespread; deepfake technology will be deployed in attempts to bypass them. Scherhag et al.[50] agree with this belief and suggest that 'face morphing' technology could be utilised by attackers to defeat modern facial recognition technology. Scherhag et al.[50] provided proof of this suggestion through a demonstration where two individual's faces were combined into a single image using a face morphing algorithm; this final image was then validated successfully. It is then suggested that by introducing an image of both the victim and attacker combined into the verification sample database, an attacker might gain access to the victim's account etc. by using their own face as verification. Scherhag, Rathgeb and Busch[49] note that the use of face morphing could allow customers at an airport to evade facial identification security systems and travel under false credentials. The author then proposes a detection method utilising the OpenFace algorithm to identify inconsistencies in facial recognition databases. However, this only reached an accuracy of approximately 83.8%. Deepfake media may also be deployed in the future to discredit political opponents, damage the reputation of companies; propagating conspiracies or misinforming voters.

During her presentation at DEF CON 27 in 2019[2], Anna Skelton was asked how organisations could correct false narratives. She responded with "Even if you can detect that a video is fake if it's already been seen; does it even matter?"

Less sophisticated deepfakes are generally referred to as 'Cheapfakes'. Cheapfakes make use of regular media editing software such as Adobe AfterEffects or Premiere[7] to mislead users across social media platforms. An infamous example is a video that circulated on Facebook in 2019, purportedly showing a high-ranking government official behaving inappropriately. This was likely done to discredit the official by making them appear unfit for their role in government. Although the video was quickly identified as fake, it had already been shared millions of times, including by prominent members of the opposition and a leading political figure.

Even without being directly deployed, deepfake technology has introduced uncertainty in situations where none previously existed. A striking example from contemporary politics occurred during an attempted coup in a nation on the 7th of January 2019. At the time, it was revealed that the country's leader had fallen seriously ill. Shortly afterward, a televised New Year address was broadcast in which the leader appeared highly unusual—blinking infrequently, displaying an abnormal posture, and lacking natural facial details like wrinkles. This led the nation's military to launch a coup, claiming that the leader had died and that the video was a deepfake fabricated by the government to maintain power[19].

Li and Lyu[38] tested a deepfake detection system they had developed against the president's conference and found that it was almost certainly a genuine video. As a result, the president likely appeared different as he had suffered a stroke and had received surgery to repair his damaged facial muscles.

In contrast to the previous articles, an expert interviewed by National Public Radio (NPR)[26] disagrees with the threat posed by deepfake-augmented misinformation. The interviewee goes on to say that deepfake disinformation is "The dog that never barked"[26] and suggests that the optimal time for deepfakes to be deployed in this manner has passed due to the public knowledge of election-meddling and disinformation. Furthermore, they continue by adding that deepfakes are not necessary to misinform those with poor technology literacy. Though this is a reasonable criticism of the threat posed by deepfake-enhanced fake news; it should be noted that the individual interviewed was taking a western-centric view, with Russia being considered the antagonist. Therefore, they were unlikely to consider the technology-literacy level of nations with less access to technology.

Ultimately, the fact that deepfake creation systems exist has created a level of doubt which may potentially poison the opinions of society, regardless of whether they have been deployed or not. Moreover, the real-world impact of the Gabon incident further establishes how important minor details are when recreating a person's likeness. Observers are quick to notice differences in an individual's face, particularly when they infrequently see them [40].

### 3.3. Countermeasures

This subsection will briefly explore methods and technologies which can be utilised to reduce the threat posed by deepfake enhanced phishing attacks.

#### 3.3.1. Video Detection

Korshunov and Marcel[34] suggest the implementation of a system that analyses the lip movements of a speaker within a video and combines them with the accompanying audio to reduce modality. It was found that a long short-term memory (LSTM)[34] model provided the best results, yet still presented an equal error rate (EER) of 14.12% during the testing stage of their project.

Korshunov and Marcel[33] further support their proposal through the creation of high-quality deepfakes and subsequently training the Support Vector Machine (SVM) classifier. This approach was found to have an EER of 8.97%. Alternative detection measures were also evaluated. These generated deepfakes were then tested against the Visual Geometry Group VGG[41] model and Facenet[51] based facial recognition models. However, these were found to be unsuitable for detecting deepfake videos. Korshunov and Marcel found that they had an EER rate of up to 95%. Finally, Korshunov and Marcel tested an LSTM system and found it to be unreliable with an EER of (41.8%).

Alongside the aforementioned methods, Li, Chang and Lyu[37] propose a system that detects the frequency at which an individual within a suspected deepfake blinks their eyes. This approach was chosen as it was been observed that deepfake videos tend to poorly emulate realistic blinking[37]. Both a long-term convolutional recurrent network (LRCN), convolutional neural network (CNN) and an 'eye aspect ratio' (EAR) trained SVM were compared. It was found that the EAR-SVM provided highly inconsistent results, with an area under receiver operating characteristic (AUROC) curve value of 0.79. In contrast, the CNN model provides 0.98 while the LRCN gives a value of 0.99.

A similar system was also proposed Yang, Li and Lyu[54] but instead, suggests that erratic head positioning can be used to identify deepfake videos by supplying an SVM classifier with facial feature bounding box values[54]. This data was found to have a range of 0.02 in real imagery while deepfake videos possessed a range of 0.08. As a result of deploying this model, testing generated an AUROC between 0.949 and 0.974

An additional approach is presented by Li and Lyu[39]. This model evaluates facial data in a similar manner to Yang, Li and Lyu[54], but instead of head positioning; artefact-induced-warping is utilised Li and Lyu[39] found that their model provided up to 99.9% accuracy when utilising their ResNet50 based CNN model against low-quality forgeries and 93.2% against the high-quality equivalent. Li and Lyu[39] noted that this model outperforms the Two-Stream NN proposed by Zhou et al[55] by 17%. Furthermore, the ResNet50 model outperforms the Headpose[54] model by approximately 8% when evaluating static imagery. As Li and Lyu's Face Warping model[39] used a different video dataset, the accuracy cannot be directly compared to the Headpose model when evaluating video data. However, as ResNet50 outperforms Headpose elsewhere it can be postulated that it would do so in other scenarios.

A comparison and summary of a variety of detection models was Produced by Sabir et al. [47] who concluded "landmark-based face alignment with bidirectional-recurrent-denset"[47] methods provided the best results in deepfake video detection. It should be noted that Sabir et al[47] do not provide the relevant statistical data to support this assertion. However, considering the comparisons made previously, this summary does appear to be correct as Li, Chang and Lyu's[37] LRCN model and the Face warping model proposed by Li and Lyu[39] provided the highest degree of accuracy. Though fully supported by data, these findings should be carefully considered, as this is not a direct comparison. As each project made use of different resolution videos; the blink evaluation model simply states that 'low quality' datasets were used, while Headpose utilised 294x500 and the ResNet50 model made use of 512x384.

Though the resultant accuracy attained is not perfect, Korshunov and Marcel[33] previously demonstrated that automated detection methods are feasible even when a human could not distinguish the validity of a video, though this requires both video and audio. Furthermore, this system may be further impacted if deepfake audio is introduced instead of the original audio. However, the Headpose[54] and Face Warping[39] models provided even more reliable results by utilising facial key points to identify inconsistencies, suggesting that methodologies utilising this approach can be considered superior for deepfake video detection. As these models are focused on multiple-frame media (videos) they do not perform as well when tested against singular images due to the lack of temporal data. Headpose, for example; produced an AUROC of 0.84 for images against 0.949 for videos.

### 3.3.2. Audio Detection

For detecting audio deepfakes; the application named Dessa[24] was recently developed. Dessa analyses audio resources Mel-frequency spectrograms (a visual interpretation of audio) for inconsistencies. This is done using a Temporal Convolution Network (TCN) in combination with a sigmoid activation function. This allows for any length of input to be processed and minimises information leakage between layers.[23]. Dessa produced a success rate of 85% when tested against a set of fake and real audio spectrograms.

PinDr0p is a commercial voice authentication system sold by Pindrop[13]. The system was initially proposed in 2010 by Balasubramaniyan et al.[17] and utilises a Linear Predictive Coding (LPC) algorithm to characterise the frequency bands of a voice. Using this audio information, Pindrop can identify the source and network path features of an incoming VoIP call with 91.6% accuracy. Pindrop may be deployed in circumstances where the home location of a caller is known, yet an attacker in a different geographic location is impersonating them. After identifying the threat of deepfake audio, Pindrop went on to develop Deep Voice 1, Deep Voice 2[15], and most recently Deep Voice 3[44]. These systems utilise the WaveNet Architecture which converts spoken language into audio frequency characteristics. DeepVoice 3 can be used to synthesise fake voices with a mean opinion score (MOS) of 95%. Vijay Balasubramaniyan, the CEO of Pindrop claims that their current technology can detect 90% of deepfake voices[18]. However, Balasubramaniyan does not provide statistics supporting this assertion.

Though there is less focus on audio deepfake detection, in favour of video detection; advances have recently been made by both Dessa and Pindrop systems. Pindrop reported its DeepVoice 3 system to be slightly more reliable with a 90% accuracy in comparison to Dessa's 85%. Though Pindrop's results are not fully available to the public due to commercial interests.

### 3.3.3. Training and Awareness

One of the most effective methods an organisation can deploy to counteract the threat of phishing attacks is user training; primarily the training of employees to identify communications that share common characteristics with phishing tactics. For example; Carella, Kotsoev and Truta[20] observed that employees provided with document-based training reduced the rate at which they erroneously interacted with phishing content by up to 42%. This resulted in only 8% of phishing attempts reaching success. The same experiment was also repeated within a classroom environment and found that phishing success dropped by 14% during the week after the class but rapidly returned to average levels in the following weeks. Melad, Basir and Saudi[14] further supports the efficacy of distributed training materials and found that embedded training (training that refers directly to systems and processes employees use) produced superior results.

Finally, the NCSC[4] agrees with the efficacy of training and additionally notes that those who prove particularly vulnerable to a targeted phishing campaign should be given further assistance and instruction.

## 4. Experimental Setting

The objective of our experiment was to present average individuals with examples of real and deepfake media to evaluate the ability of the public to correctly identify deepfake phishing attacks. To this end, we created deepfake video and audio and concealed them among genuine examples. We then presented them to the participants of our experiments as described in this section.

### 4.1. Hardware Specifications

The deepfakes produced in this research utilised a low-spec computing facility, a personal desktop machine. The following details the specifications of the machine. Operating System: Windows 10 Home 64-bit (10.0 build 18362); Processor: Intel i7-8700K 3.70 GHz (12 Cores); GPU: NVIDIA GeForce GTX 1080Ti 11GB; RAM: Corsair Vengeance DDR4 32GB.

## 4.2. Software Deployed

**DeepFaceLab** DeepFaceLab is an open-source video deepfake creation software developed in 2018 by 'iperov' (Ivan Perov). It utilises a generative adversarial network (GAN) and is based on the Python coding language and Tensor-Flow platform. The DeepfaceLab project also aims to implement a "state-of-the-art framework" and provide a "simple and flexible model construction"[42] through the use of 'Leras' (Lighter Keras). This is mirrored by its highly modular construction that allows the user to experiment with and develop a custom methodology through sequential implementation. On its Github repository, DeepFaceLab claims that "more than 95% of deepfake videos are created with DeepFaceLab"[28].

**Resemble AI** Resemble is a commercial application designed to create 'characters' through the implementation of audio deepfake technology. The software was launched in 2019 by a company of the same name to enable the creation of realistic and emotive artificial voices within the entertainment industry[8]. Furthermore, the API is also made available to aid integration with customers' existing equipment. Resemble has also released an open-source software named 'Resemblyzer'[12].

**Speech-Driven Animation** Speech-Driven Animation[9] makes use of a temporal GAN with two classifiers to develop deepfake lip-synching when provided with sample imagery and audio. Speech-Driven Animation locates the key features of the entire subject's face and then uses this gathered data to synthesise a face synchronised with the input audio data. Each of the application's two discriminators fulfils a different purpose. The first aims to maximise the realism of a singular frame while the second's objective is to ensure the realistic transition between frames in the sequence[52].

## 4.3. Dataset Description

The dataset utilised to create the video component was supplied by Conrad Sanderson[1] and the associated academic publication[48]. This dataset consists of video and audio recordings of 43 individuals. These individuals are situated in an identical setting and repeat ten sentences, two are the same across all actors while the rest were selected at random. These recordings take place in multiple sessions. With the first six sentences recited in session one while two further sentences were recorded in both sessions two and three. Furthermore, each actor provided a "head rotation sequence" within each session. The video components have a resolution of 512 x 384 and are stored as thirteen folders of JPEG images; numbered sequentially. The audio components are also stored sequentially and make use of a single sound channel, 16-bit audio depth and a frequency of 32kHz. These are stored as Wav files. The dataset was downloaded as 43 separate zip files, each consisting of 71Mb of data per actor, and utilised in the creation of audio and video in our experiments.

## 4.4. Questionnaire Design

The questionnaire was split into three primary sections. The first aimed to identify the accuracy at which participants may detect video deepfakes, while the second focused on audio deepfakes. The final section ascertained the accuracy with which participants identified both audio and video deepfakes when they were presented in an integrated manner; while also being combined through the use of an artificial lip-synching application.

By presenting the questionnaire in this way, the realism of the video, audio elements could be assessed independently, followed by media in which they are combined. This also has the advantage of mirroring real-world scenarios, where a deepfake-enabled phishing attack is not guaranteed to utilise multiple elements.

### 4.4.1. Evaluation Characteristics

Participants were presented with a combination of genuine and deepfake material in a randomised order. The composition of these videos were as follows:

- Video evaluation: 8 videos total, with 3 fake and 5 genuine.

- Audio evaluation: 13 audio clips total with 3 fake and 10 genuine.

- Combined evaluation: 8 clips total with a varied combination of real and genuine audio and video combined through the use of a deepfake lip-synch application.

By presenting the videos in such a way, the respondents were required to critically analyse the media presented and attempt to pick the fake examples out themselves instead of being asked to numerically quantify how realistic they believed a deepfake to be.

### 4.4.2. Evaluation Length

The evaluation material provided was kept brief in length for three primary reasons. The first was to emulate a real-world situation in which an attacker would aim to minimise the victim's opportunity to analyse the fake media. Secondly, producing a lengthy deepfake with a high degree of realism requires a great deal of time, not only during the training and generation of a video etc. but also during the editing process in which volume and errors in the algorithm may be adjusted or removed. Finally, if presented with three videos each of five minutes, it is unlikely the questionnaire respondents would be engaged throughout the entire evaluation and instead would begin to rush their analysis or simply give up during the process.

### 4.4.3. Questions

First, participants were asked to select the age bracket they belonged to. This allowed for evaluation data to be sorted by age group. Question two and the final question asked the participants to report how much of a threat they believed deepfake media to be (or if they were previously unaware of its existence). These responses were then contrasted to evaluate whether participants' opinions changed after exposure to deepfake media (opinion before and after). The rest of the questions were aimed at evaluating deepfake video and audio created in our experiments. Respondents were asked to identify the examples they believed to be deepfakes out of a selection clips. They were then asked to justify why they believed the clips they selected were fake. These questions were repeated three times; once for each of the types of deepfake (video, audio and combined). By asking these questions it could be observed how statistically likely an individual is to correctly or incorrectly identify a deepfake. Furthermore, by asking for the participants' reasoning behind their selection, the weaknesses of the created media could be identified.

Two identical questionnaires (form A and B) were utilised. Both contain the same questions and deepfake content. However, form A presented to participants who are familiar with the author; particularly the author's voice. Form B presented to those who are not familiar with the author (Friends of friends etc.). This is done as the author's voice is to be used throughout the audio section and combined section of the questionnaire and as a result; the author's friends and family may immediately identify deepfakes where it is used in conjunction with another actor's likeness. Therefore, by segregating these responses, they will not statistically impact the responses gathered by form B.

#### 4.4.4. Deepfake Examples

This section presents a sample (selected frames) of deepfake videos created by the authors and presented to participants in our experiments. Due to space constraints only three selected samples are presented (see Figures 1, and 2).

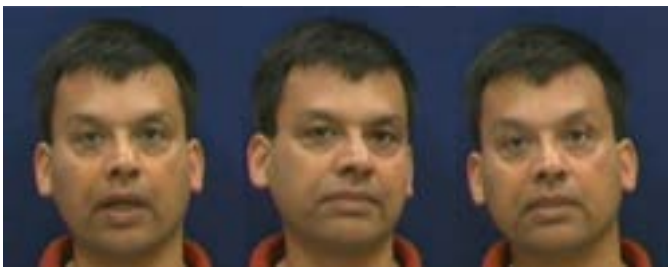

**Figure 1:** Example deepfake produced and presented to survey participants by authors, combining a pair of actors from Sanderson's dataset [1].

Figure 2 provides multiple frames generated through the use of the Speech-Driven Animation[9] application's 'Timit' model. These outputs were found to be of low quality as the program only provides a low-resolution output.

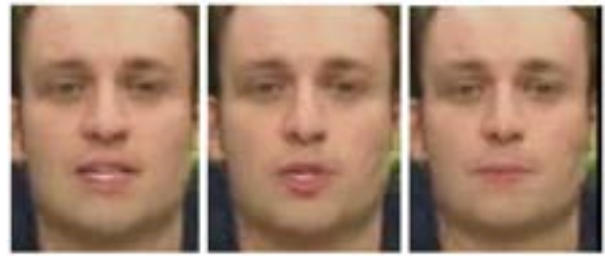

**Figure 2:** Multiple frames generated by authors through the use of the Speech-Driven Animation[9] application's 'Timit' model.

**Table 1**
Respondent data sorted by self-identified age bracket.

| Age Range | Participants | Percentage of Participants |
|---|---|---|
| Below 18 | 1 | 2.27% |
| 18 - 24 | 11 | 25.00% |
| 25 - 34 | 10 | 22.73% |
| 35 - 44 | 4 | 9.09% |
| 45 - 54 | 7 | 15.91% |
| 55 - 64 | 5 | 11.36% |
| 65 - 74 | 4 | 9.09% |
| 75 and above | 2 | 4.55% |
| Total | 44 | 100.00% |

## 5. Results and Discussion

Forty-four individuals participated in the assessment of spearphishing-like attack scenarios, i.e. artificial audio and video media created using deepfake technology in our experiment. The age distribution of these respondents is shown in Table 1.

It was found that over 90% of respondents believed that deepfake media posed a threat to society, with 44% believing it to be a severe threat while a further 48% felt it was a moderate threat. Furthermore, after evaluating the media presented to them 52.7% felt that artificial media was more of a threat than they first thought while 45% felt their initial analysis was correct (Figure 3).

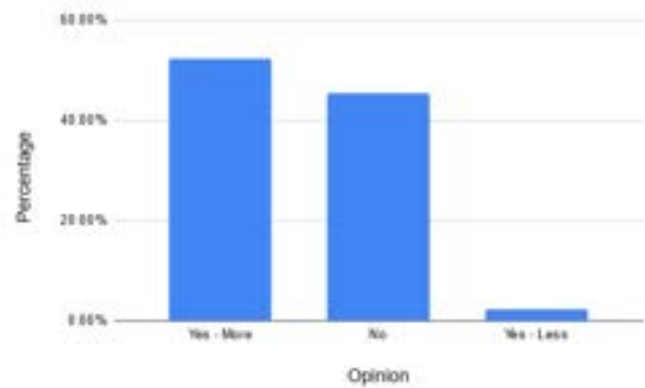

**Figure 3:** Distribution of participants' revised opinions on the threat of deepfakes.

**Table 2**
Video and Audio identification rate by age.

| Age range | Video Detection Rate | Audio Detection Rate |
|---|---|---|
| Below 18 | 33.30% | 66.70% |
| 18 - 24 | 45.30% | 56.10% |
| 25 - 34 | 68.30% | 45.80% |
| 35 - 44 | 17.50% | 56.30% |
| 45 - 54 | 21.40% | 47.60% |
| 55 - 64 | 9.50% | 28.80% |
| 65 - 74 | 30.80% | 47.90% |
| Above 75 | 0.00% | 25% |

**Table 3**
Form A Combined deepfake component detection rates by age range.

| Age Range | Video Detection Rate | Audio Detection Rate |
|---|---|---|
| Below 18 | N/A | N/A |
| 18 - 24 | 76.00% | 38.70% |
| 25 - 34 | 100.00% | 100.00% |
| 35 - 44 | 100.00% | 0.00% |
| 45 - 54 | 72.50% | 23.80% |
| 55 - 64 | 61.10% | 25.00% |
| 65 - 74 | N/A | N/A |
| Above 75 | N/A | N/A |

Audio deepfakes were found to be the most difficult to successfully detect when compared to video deepfake detection rates (Table 2). This presents malicious individuals with an opportunity to use the technology to carry out successful phishing attacks. Using this information it can be seen that of the audio selections made 66.2% were incorrect while 33.8% were correct, compared to 56.9% correct and 43.1% incorrect when selecting video examples. This data is compared in Figure 4 and Figure 5.

**Table 4**
Form B Combined deepfake component detection rates by age range.

| Age Range | Video Detection Rate | Audio Detection Rate |
|---|---|---|
| Below 18 | 50.00% | 0.00% |
| 18 - 24 | 91.70% | 38.90% |
| 25 - 34 | 70.40% | 13.00% |
| 35 - 44 | 66.70% | 61.10% |
| 45 - 54 | 41.30% | 33.30% |
| 55 - 64 | 57.10% | 21.40% |
| 65 - 74 | 50.00% | 30.20% |
| Above 75 | 81.90% | 8.30% |

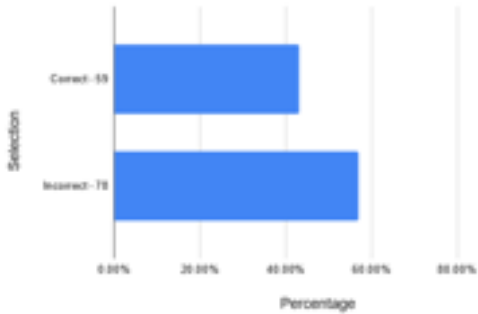

**Figure 4:** Comparing correct and incorrect selections when attempting to identify fake videos.

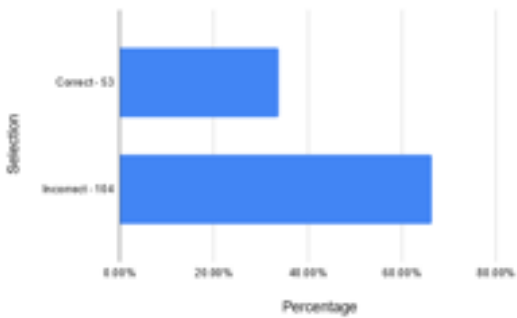

**Figure 5:** Comparing correct and incorrect selections when attempting to identify fake audio.

Interestingly, when audio and video components are combined the rate at which individuals detect fake audio remains nearly the same (68.3% incorrect versus 31.7% correct) while video aspects become easier to detect (30.2% incorrect versus 69.8% correct). This information is presented in Table 3 and Table 4. This may be due to the poor quality of the lip-synching application's output. Furthermore, it could be suggested that when an individual believed something was 'off' but was not fully sure what, they may default to selecting the video option as this is the most obvious component (Figure 6 and Figure 7).

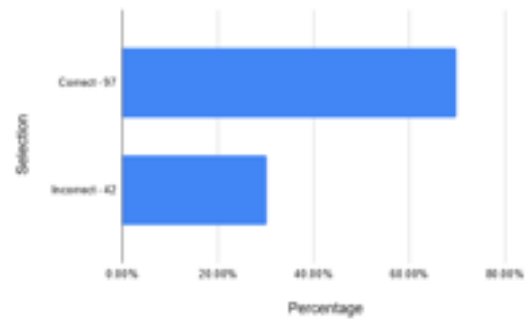

**Figure 6:** Comparing correct and incorrect selections when attempting to identify fake video components.

When comparing the average successful detection rates between those that were previously aware of deepfakes and those that were not, it was found that those who had prior knowledge of the technology had a success rate of 42%. However, those who did not have this prior knowledge had a 22% success rate. Figure 8 compares these success rates.

It was also found that there was little difference between those that would recognise the author's voice and those that would not. The only exception to this is among the 25 – 34-year-old age group when evaluating combined deepfakes. In this exception, as can be seen in Figure 9 and Figure 10 the participant within the age bracket who knew the author's voice did not misidentify any real audio segments (Though once again, the sample size provided by Form A limits the value of this data). In comparison to those who were given Form B who produced a success rate of 13% due to a high rate of false positives (Figure 9 and Figure 10).

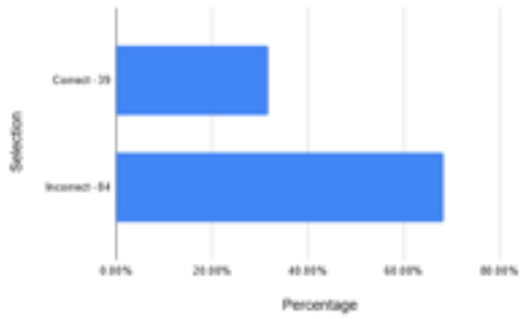

**Figure 7:** Comparing correct and incorrect selections when attempting to identify fake audio components.

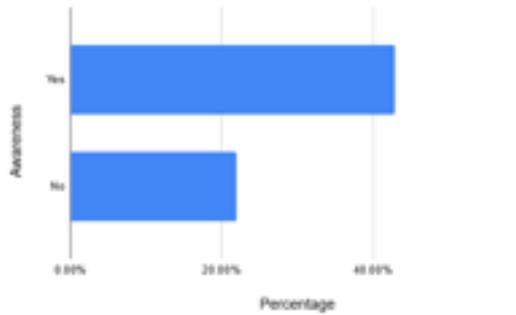

**Figure 8:** A comparison of detection rates between those with knowledge of deepfakes and those without.

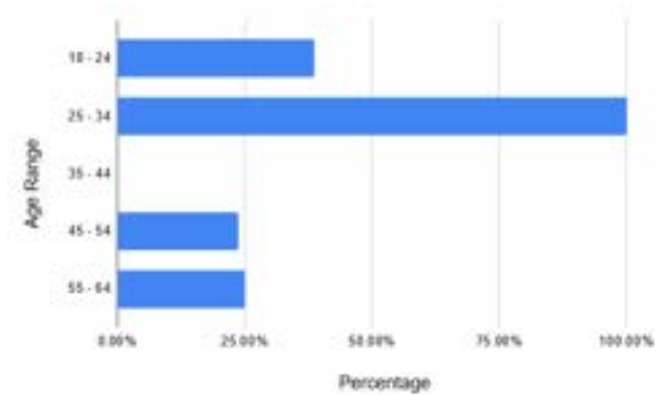

**Figure 9:** Comparing success rates across ages (Form A Combined Audio).

Additionally, it was found that though a person may not be able to identify exactly what it is about a video they believe makes it appear fake; it can be seen that participants still perceived as they said "something off" or "unnatural". This might be a subconscious 'uncanny valley' effect where something appears very nearly human but instils a sense of unease in its observer due to it being quite authentic but not sufficiently so.

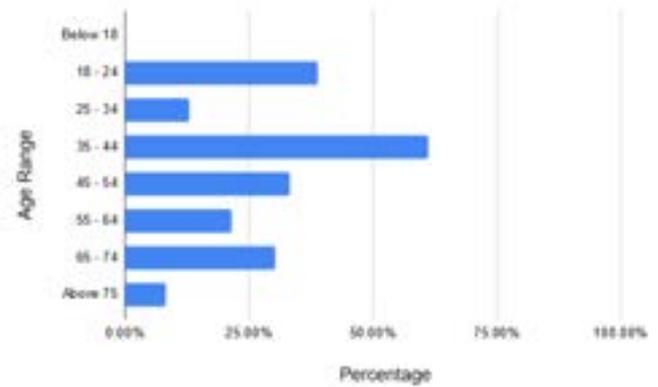

**Figure 10:** Comparing success rates across ages (Form B Combined Audio).

Finally, though not fully explored, it could be theorised that the high percentage of false positives provided by the participants was a side effect of their knowledge that one or more deepfakes were present. This uncertainty, therefore, increasing the chance of believing genuine videos to be doctored in some way. This effect can also be seen in the aforementioned case in Gabon[19] where a coup was nearly incited due to the knowledge that deepfake technology existed.

## 6. Conclusion and Future Works

It can ultimately be concluded that deepfake-enabled phishing attacks pose a threat to modern society. For example, if only 34% of assumptions were correct when attempting to identify fake audio samples then it may be assumed that the remaining 66% of attacks would convince a victim into falling for a phishing attack. This deduction has real-world merit as can be seen in the previously discussed case[10]. Furthermore, it was found that creating deepfake media is exceptionally easy with only moderate research required into how to operate the required applications and little need for knowledge of the underlying technologies. Our experiments demonstrate that creating these types of attacks can be done using a low-spec computing facility such as a desktop or laptop computer. This further enforces the conclusion that these types of attacks will be more commonly deployed in the coming years.

Based on the findings produced by the evaluation of the deepfake media, multiple further conclusions can be reached when identifying whether there are specific demographics particularly vulnerable to deepfake-enabled attacks. For example, it was found that the older an individual is, the less likely they are to correctly identify artificial media; resulting in a greater chance of the elderly being victimised. Moreover, those who are unfamiliar with artificially generated media are 20% less accurate when attempting to identify it. In addition to this, it can be concluded that deepfake audio com-ponents appear to be the most difficult to identify in comparison to video examples. Furthermore, there appears to be little difference between those who recognise a voice and those who don't when considering deepfake audio samples.

Finally, it can be concluded that the use of lip-synch software greatly increases the rate at which video deepfakes are detected while having a negligible effect on artificial audio detection rates. This is likely due to the poor output quality provided by existing technology suggesting that this specific technology is yet to advance sufficiently. The lip-synching situation may potentially be circumnavigated by having a destination actor speaking the lines themselves before performing the deepfake process as this may preserve the original moth movements. However, this would require further investigation. In future, we plan to use tools like Maltego[11] to automate the retrieval of one's available data (images, videos and audio) from social media and use them to create similar spearphishing attacks. Such a tool can be useful in security awareness programs and for ethical hacking purposes.

## Compliance with Ethical Standards

The ethical considerations of this research were evaluated by Robert Gordon University in the UK and was subsequently approved. (SPER form ID 130962886).

## References

[1] , 2008. Vidtimit audio-video dataset. URL: http://conradsanderson.id.au/vidtimit/.

[2] , 2019. Anna skelton - analyzing the effects of deepfakes on market manipulation - def con 27 ai village. URL: https://www.youtube.com/watch?v=BlS-inTxnp8.

[3] , 2019. Conmen made €8m by impersonating french minister - israeli police. URL: https://www.theguardian.com/world/2019/mar/28/conmen-made-8m-by-impersonating-french-minister-israeli-police.

[4] , 2019. Phishing attacks: defending your organisation. URL: https://www.ncsc.gov.uk/guidance/phishing.

[5] , 2019. Why deepfakes pose an unprecedented threat to businesses. URL: https://aibusiness.com/document.asp?doc%5C_id=760904.

[6] , 2020. URL: https://publish.twitter.com/?query=https%3A%2F%2Ftwitter.com%2FElonmusk%2Fstatus%2F1256239815256797184&widget=Tweet.

[7] , 2020. Adobe. vfx amp; motion graphics software | adobe after effects. URL: https://www.adobe.com/uk/products/aftereffects.html.

[8] , 2020. Ai generated voices  resemble ai. URL: https://www.resemble.ai/.

[9] , 2020. Dinoman/speech-driven-animation. URL: https://github.com/DinoMan/speech-driven-animation.

[10] , 2020. Fraudsters used ai to mimic ceo's voice in unusual cybercrime case. URL: https://www.wsj.com/articles/fraudsters-use-ai-to-mimic-ceos-voice-in-unusual-cybercrime-case-11567157402.

[11] , 2020. Maltego. URL: https://www.maltego.com/.

[12] , 2020. resemble-ai/resemblyzer. URL: https://github.com/resemble-ai/Resemblyzer.

[13] , 2020. Voice biometric authentication amp; anti-fraud for call centers. URL: https://www.pindrop.com/.

[14] Al-Daeef, M.M., Basir, N., Saudi, M.M., 2017. Security awareness training: A review, in: Proceedings of the World Congress on Engineering, pp. 5–7.

[15] Arik, S.O., Chrzanowski, M., Coates, A., Diamos, G., Gibiansky, A., Kang, Y., Li, X., Miller, J., Ng, A., Raiman, J., et al., 2017. Deep voice: Real-time neural text-to-speech. arXiv preprint arXiv:1702.07825 .

[16] Babuta, A., Oswald, M., Janjeva, A., 2020. Artificial intelligence and uk national security.

[17] Balasubramaniyan, V.A., Poonawalla, A., Ahamad, M., Hunter, M.T., Traynor, P., 2010. Pindr0p: using single-ended audio features to determine call provenance, in: Proceedings of the 17th ACM conference on Communications and computer security, pp. 109–120.

[18] Burt, C., 2019. Threat of deepfakes draws legislator and biometrics industry attention. URL: https://www.biometricupdate.com/201902/threat-of-deepfakes-draws-legislator-and-biometrics-industry-attention.

[19] Cahlan, S., 2020. Analysis | how misinformation helped spark an attempted coup in gabon. URL: https://www.washingtonpost.com/politics/2020/02/13/how-sick-president-suspect-video-helped-sparked-an-attempted-coup-%gabon/.

[20] Carella, A., Kotsoev, M., Truta, T.M., 2017. Impact of security awareness training on phishing click-through rates, in: 2017 IEEE International Conference on Big Data (Big Data), IEEE. pp. 4458–4466.

[21] Centre, N.C.S., 2020. URL: https://www.ncsc.gov.uk/guidance/whaling-how-it-works-and-what-your-organisation-can-do-about-it.

[22] for Data Ethics, C., Innovation, 2019. Snapshot Paper - Deepfakes and Audiovisual Disinformation.

[23] Dessa, 2019. Detecting audio deep fakes with ai. URL: https://medium.com/dessa-news/detecting-audio-deepfakes-f2edfd8e2b35.

[24] Dessa-Oss, 2020. dessa-oss/fake-voice-detection. URL: https://github.com/dessa-oss/fake-voice-detection.

[25] Department for Digital, Culture, M., Sport, 2020. Cyber Security Breaches Survey. Ph.D. thesis.

[26] Ewing, P., 2020. Why fake video, audio may not be as powerful in spreading disinformation as feared. URL: https://www.npr.org/2020/05/07/851689645/why-fake-video-audio-may-not-be-as-powerful-in-spreading-%disinformation-as-feare?t=1595494301688.

[27] Forbes, 2020. Why deepfakes are a net positive for humanity. URL: https://www.forbes.com/sites/simonchandler/2020/03/09/why-deepfakes-are-a-net-positive-for-humanity/#e512a3b2f84f.

[28] Iperov, 2020. iperov/deepfacelab. URL: https://github.com/iperov/DeepFaceLab.

[29] Jia, Y., Zhang, Y., Weiss, R., Wang, Q., Shen, J., Ren, F., Nguyen, P., Pang, R., Moreno, I.L., Wu, Y., et al., 2018. Transfer learning from speaker verification to multispeaker text-to-speech synthesis, in: Advances in neural information processing systems, pp. 4480–4490.

[30] Karnouskos, S., 2020. Artificial intelligence in digital media: The era of deepfakes. IEEE Transactions on Technology and Society 1, 138–147.

[31] Karras, T., Laine, S., Aittala, M., Hellsten, J., Lehtinen, J., Aila, T., 2020. Analyzing and improving the image quality of stylegan, in: Proceedings of the IEEE/CVF Conference on Computer Vision and Pattern Recognition, pp. 8110–8119.

[32] Klahr, R., 2017. Cyber security breaches survey. Ph.D. thesis. University of Portsmouth.

[33] Korshunov, P., Marcel, S., 2018a. Deepfakes: a new threat to face recognition? assessment and detection. arXiv preprint arXiv:1812.08685 .

[34] Korshunov, P., Marcel, S., 2018b. Speaker inconsistency detection in tampered video, in: 2018 26th European Signal Processing Conference (EUSIPCO), IEEE. pp. 2375–2379.

[35] Langlois, S., 2018. This day in history: Hacked ap tweet about white house explosions triggers panic. URL: https://www.marketwatch.com/story/this-day-in-history-hacked-ap-tweet-about-white-house-explosions-%triggers-panic-2018-04-23.

[36] Le, B.H., Zhu, M., Deng, Z., 2013. Marker optimization for facial motion acquisition and deformation. IEEE transactions on visualization and computer graphics 19, 1859–1871.

[37] Li, Y., Chang, M.C., Lyu, S., 2018. In ictu oculi: Exposing ai generated fake face videos by detecting eye blinking. arXiv preprint

arXiv:1806.02877 .

[38] Li, Y., Lyu, S., 2018a. Exposing deepfake videos by detecting face warping artifacts. arXiv preprint arXiv:1811.00656 .

[39] Li, Y., Lyu, S., 2018b. Exposing deepfake videos by detecting face warping artifacts. arXiv preprint arXiv:1811.00656 .

[40] Liu, Z., Zhang, Z., 2011. Face geometry and appearance modeling: concepts and applications. Cambridge University Press.

[41] Parkhi, O.M., Vedaldi, A., Zisserman, A., 2015. Deep face recognition .

[42] Petrov, I., Gao, D., Chervoniy, N., Liu, K., Marangonda, S., Umé, C., Jiang, J., RP, L., Zhang, S., Wu, P., et al., 2020. Deepfacelab: A simple, flexible and extensible face swapping framework. arXiv preprint arXiv:2005.05535 .

[43] Phishing.Org, . History of phishing. URL: https://www.phishing.org/history-of-phishing.

[44] Ping, W., Peng, K., Gibiansky, A., Arik, S.O., Kannan, A., Narang, S., Raiman, J., Miller, J., 2017. Deep voice 3: Scaling text-to-speech with convolutional sequence learning. arXiv preprint arXiv:1710.07654 .

[45] Rader, M., Rahman, S., 2015. Exploring historical and emerging phishing techniques and mitigating the associated security risks. arXiv preprint arXiv:1512.00082 .

[46] Rekouche, K., 2011. Early phishing. arXiv preprint arXiv:1106.4692 .

[47] Sabir, E., Cheng, J., Jaiswal, A., AbdAlmageed, W., Masi, I., Natarajan, P., 2019. Recurrent convolutional strategies for face manipulation detection in videos. Interfaces (GUI) 3.

[48] Sanderson, C., Lovell, B.C., 2009. Multi-region probabilistic histograms for robust and scalable identity inference, in: International conference on biometrics, Springer. pp. 199–208.

[49] Scherhag, U., Rathgeb, C., Busch, C., 2018. Towards detection of morphed face images in electronic travel documents, in: 2018 13th IAPR International Workshop on Document Analysis Systems (DAS), IEEE. pp. 187–192.

[50] Scherhag, U., Rathgeb, C., Merkle, J., Breithaupt, R., Busch, C., 2019. Face recognition systems under morphing attacks: A survey. IEEE Access 7, 23012–23026.

[51] Schroff, F., Kalenichenko, D., Philbin, J., 2015. Facenet: A unified embedding for face recognition and clustering, in: Proceedings of the IEEE conference on computer vision and pattern recognition, pp. 815–823.

[52] Vougioukas, K., Petridis, S., Pantic, M., 2018. End-to-end speech-driven facial animation with temporal gans. arXiv preprint arXiv:1805.09313 .

[53] Wojewidka, J., 2020. The deepfake threat to face biometrics. Biometric Technology Today 2020, 5–7.

[54] Yang, X., Li, Y., Lyu, S., 2019. Exposing deep fakes using inconsistent head poses, in: ICASSP 2019-2019 IEEE International Conference on Acoustics, Speech and Signal Processing (ICASSP), IEEE. pp. 8261–8265.

[55] Zhou, P., Han, X., Morariu, V.I., Davis, L.S., 2017. Two-stream neural networks for tampered face detection, in: 2017 IEEE Conference on Computer Vision and Pattern Recognition Workshops (CVPRW), IEEE. pp. 1831–1839.

[56] Zhu, B., Fang, H., Sui, Y., Li, L., 2020. Deepfakes for medical video de-identification: Privacy protection and diagnostic information preservation, in: Proceedings of the AAAI/ACM Conference on AI, Ethics, and Society, pp. 414–420.